\documentclass[prd,showpacs,showkeys,nofootinbib,twocolumn,superscriptaddress]{revtex4-2}

\usepackage{graphicx}
\usepackage{amsmath,amsfonts,amssymb,amsxtra,latexsym}

\usepackage{hyperref}
\hypersetup{
    colorlinks=true,
    linkcolor=red,
    citecolor=blue,      
    urlcolor=blue,
}

\usepackage[utf8]{inputenc}

\begin{document}

\title{Colliding Plane Fronted Waves and a Gravito--Electromagnetic Searchlight}

\author{Peter A. Hogan}
\email{peter.hogan@ucd.ie}
\affiliation{School of Physics, University College Dublin, Belfield, Dublin 4, Ireland} 

\author{Dirk Puetzfeld}
\email{dirk@puetzfeld.org}
\homepage{http://puetzfeld.org}
\affiliation{School of Physics, University College Dublin, Belfield, Dublin 4, Ireland}
\affiliation{University of Bremen, Center of Applied Space Technology and Microgravity (ZARM), 28359 Bremen, Germany} 

\date{ \today}

\begin{abstract}
We present a formulation of Einstein--Maxwell vacuum fields due to plane fronted electromagnetic waves sharing their wave fronts with gravitational waves. This is based on a recent geometrical reconstruction of plane fronted wave fields by the authors which clearly identifies the cases in which the wave fronts collide or do not collide. In the former case our construction suggests an explicit example of a searchlight beam, accompanied by gravitational radiation, which sweeps across the sky. This gravito--electromagnetic searchlight and its properties are described in detail.
\end{abstract}

\pacs{04.20.-q; 04.20.Jb; 04.20.Cv}
\keywords{Classical general relativity; Exact solutions; Fundamental problems and general formalism}

\maketitle


\section{Introduction}\label{sec:1}

We describe a formulation of Einstein--Maxwell vacuum fields of plane fronted electromagnetic waves sharing their wave fronts with gravitational waves. This follows our recent reconstruction \cite{Hogan:Puetzfeld:2021:1,Hogan:Puetzfeld:2022:book} of plane fronted waves which enables a clear distinction between plane waves with colliding wave fronts (Kundt \cite{Kundt:1961} waves) and waves with non--colliding wave fronts ($pp$--waves \cite{Ehlers:Kundt:1962,Stephani:etal:2003:book}). Our new formulation suggests a simple example for which the gravitational and electromagnetic radiation fields vanish if the waves are \emph{not} colliding. This is an example of a searchlight beam, accompanied by gravitational radiation, sweeping across the sky. We refer to it as a \emph{gravito--electromagnetic searchlight}. Such waves are the asymptotic limit of spherical waves emitted by an isolated source \cite{Hogan:Puetzfeld:2021:2}. The wave fronts are colliding or not colliding depending upon the motion of the source. The paper is organized as follows: in section \ref{sec:2} we give a detailed description of our formulation of plane fronted waves in the context of Einstein--Maxwell theory. This is followed in section \ref{sec:3} by the gravito--electromagnetic searchlight which is an explicit solution of the Einstein--Maxwell equations which exploits variables which arise naturally in our geometrical construction of plane fronted waves. The solution is also described in coordinates closely associated with rectangular Cartesians and time as this makes the solution more surveyable. This leads to further properties of the solution described in section \ref{sec:4} by making use of tensor fields on a background Minkowskian space time. The paper ends with a discussion of our results in section \ref{sec:5}.

\section{Einstein--Maxwell Vacuum Fields}\label{sec:2}

We use units for which the gravitational constant $G=1$ and the speed of light in a vacuum $c=1$ and we choose the sign conventions of Synge \cite{Synge:1960:book}. Thus in a local coordinate system $\{x^i\}$ we obtain from the Riemann curvature tensor components $R_{ijkm}$ the Ricci tensor components $R_{jk}=g^{im}\,R_{ijkm}$. The Ricci scalar $R=g^{jk}\,R_{jk}$, the Einstein tensor components $G_{jk}=R_{jk}-\frac{1}{2}\,g_{jk}\,R$ and the Einstein--Maxwell vacuum field equations
\begin{equation}\label{1}
G_{jk}=-\kappa\,E_{jk}\ \ \ {\rm with}\ \ \ E_{jk}=F_{jp}\,F_{k}{}^p-\frac{1}{4}\,g_{jk}\,F_{mp}\,F^{mp}\ ,
\end{equation}
with the Maxwell tensor $F_{ij}=-F_{ji}$ satisfying Maxwell's vacuum field equations
\begin{equation}\label{2}
F^{ij}{}_{;j}=0\ \ \ {\rm with}\ \ \ F_{ij;k}+F_{ki;j}+F_{jk;i}=0\ .
\end{equation}
In (\ref{1}) $\kappa=8\,\pi$, $E_{jk}=E_{kj}$ with $E^j{}_j=0$ is the electromagnetic energy--momentum tensor, indices are raised with $g^{ij}$ and lowered with $g_{ij}$ and $g^{ij}$ is defined by $g^{ij}\,g_{jk}=\delta^i_k$ and $g_{ij}=g_{ji}$ are the components of the metric tensor. The semicolon denotes covariant differentiation with respect to the Riemannian connection calculated with the metric tensor $g_{ij}$. We shall denote by $A^i$ the components of a 4--potential from which the components of the Maxwell tensor can be obtained via
\begin{equation}\label{3}
F_{ij}=A_{j;i}-A_{i;j}=A_{j,i}-A_{i,j}\ ,
\end{equation}
with the comma denoting partial differentiation with respect to the coordinates $x^i$. With $F_{ij}$ given by (\ref{3}) the second equation in (\ref{2}) is satisfied. The first equation in (\ref{2}) can be written in terms of the 4--potential as (\cite{Synge:1960:book}, p.\ 357)
\begin{equation}\label{4}
g^{jk}\,A_{i;j;k}+R_{ij}\,A^j=0\ \ \ {\rm provided}\ \ \ A^i{}_{;i}=0\ .
\end{equation}
This relies on the Ricci identities in Synge's form
\begin{equation}\label{5}
A_{i;j;k}-A_{i;k;j}=A_p\,R^p{}_{ijk}\ ,
\end{equation}
which reveals his convention for the definition of the Riemann curvature tensor. Making use of this, following the substitution of (\ref{3}) into the first of Maxwell's equations (\ref{2}), results in (\ref{4}). As a final preliminary we note the components of the Weyl conformal curvature tensor $C_{ijkm}$ are given by
\begin{eqnarray}
C_{ijkm}&=&R_{ijkm}+\frac{1}{2}\Big(g_{ik}\,R_{jm}+g_{jm}\,R_{ik} \nonumber \\
&&-g_{im}\,R_{jk}-g_{jk}\,R_{im}\Big)\nonumber\\
&&+\frac{1}{6}\,R\,(g_{im}\,g_{jk}-g_{ik}\,g_{jm})\ .\label{6}
\end{eqnarray}

We now consider a novel form for the line element of the space--time model for plane fronted gravitational waves sharing their wave fronts with electromagnetic waves which has been constructed in \cite{Hogan:Puetzfeld:2021:1,Hogan:Puetzfeld:2022:book}:
\begin{equation}\label{7}
ds^2=g_{ij}\,dx^i\,dx^j=2\,|d\zeta-\beta(u)\,v\,du|^2+2\,q\,du\,(dv+H\,du)\ ,
\end{equation}
with
\begin{equation}\label{8}
q(\zeta, \bar\zeta, u)=\beta(u)\,\bar\zeta+\bar\beta(u)\,\zeta+\frac{1}{2}\,\gamma(u)\ .
\end{equation}
The coordinate $\zeta$ is complex with complex conjugate $\bar\zeta$ while $u, v$ are two real coordinates. The function $\beta(u)$ is an arbitrary complex--valued function of the coordinate $u$ with complex conjugate denoted by $\bar\beta(u)$, $\gamma(u)$ is an arbitrary real--valued function of $u$, and $H(\zeta, \bar\zeta, u)$ is a real--valued function to be determined by the field equations. The hypersurfaces $u={\rm constant}$ are null and generated by the shear--free, expansion--free, geodesic integral curves of the null vector field $\partial/\partial v$. It is demonstrated explicitly in \cite{Hogan:Puetzfeld:2021:1,Hogan:Puetzfeld:2022:book} that these null hypersurfaces intersect, and thus the wave fronts collide, if and only if $\beta(u)\neq 0$. Labelling the coordinates $x^i=(\zeta, \bar\zeta, v, u)$ for $i=1,2, 3, 4$ we find that the only non--vanishing components of the Riemann curvature tensor are
\begin{equation}\label{9}
R_{1424}=-q\,H_{\zeta\bar\zeta}\ ,\ R_{1414}=-q\,H_{\zeta\zeta}\ ,\ R_{2424}=-q\,H_{\bar\zeta\bar\zeta}\ ,
\end{equation}
with the subscripts here, and in the sequel, denoting partial derivatives. The Ricci tensor has only one non--vanishing component
\begin{equation}\label{10}
R_{44}=2\,q\,H_{\zeta\bar\zeta}\ .
\end{equation}
Consequently the Ricci scalar vanishes and the non--vanishing components of the Weyl conformal curvature tensor are
\begin{equation}\label{11}
C_{1414}=-q\,H_{\zeta\zeta}\ \ \ {\rm and}\ \ \ C_{2424}=-q\,H_{\bar\zeta\bar\zeta}\ .
\end{equation}
Next we consider a 4--potential on the space--time with line element (\ref{7}) given by the 1--form
\begin{equation}\label{12}
A=f(\zeta, \bar\zeta, u)\,du\ ,
\end{equation}
with $f$ a real--valued function of its arguments. This leads to a candidate for Maxwell field $F_{ij}$ having non--vanishing components
\begin{equation}\label{13}
F_{14}=f_{\zeta}\ \ \ {\rm and}\ \ \ F_{24}=f_{\bar\zeta}\ ,
\end{equation}
with the subscripts denoting partial derivatives as before. Now Maxwell's vacuum field equations (\ref{2}) are satisfied provided $f$ satisfies
\begin{equation}\label{14}
f_{\zeta\bar\zeta}=0\ ,
\end{equation}
and thus $f_{\zeta}$ is an analytic function of $\zeta$ with an arbitrary dependence on $u$. The electromagnetic energy--momentum tensor in (\ref{1}) has one non--vanishing component
\begin{equation}\label{15}
E_{44}=2\,f_{\zeta}\,f_{\bar\zeta}\ .
\end{equation}
With (\ref{10}) and (\ref{15}) the Einstein--Maxwell field equations (\ref{1}) provide the following single equation
\begin{equation}\label{16}
H_{\zeta\bar\zeta}=-\kappa\,q^{-1}f_{\zeta}\,f_{\bar\zeta}\ .
\end{equation}
In order to survey what we have here it is useful to introduce the null tetrad $k^i, l^i, m^i, \bar m^i$ given in the coordinates $x^i=(\zeta, \bar\zeta, v,u)$ for $i=1, 2, 3, 4$ by
\begin{eqnarray}
k^i&=&\delta^i_3\ ,\quad m^i=\delta^i_2\ ,\quad \bar m^i=\delta^i_1\ , \nonumber \\
l^i&=& q^{-1}\beta\,v\,\delta^i_1+q^{-1}\bar\beta\,v\,\delta^i_2-q^{-1}H\,\delta^i_3+q^{-1}\delta^i_4\ ,\ \label{17}
\end{eqnarray}
and the three complex bivectors
\begin{eqnarray}
L^{ij}&=&\bar m^i\,l^j-\bar m^j\,l^i\ ,\label{18}\\
M^{ij}&=&m^i\,\bar m^j-\bar m^i\,m^j+k^i\,l^j-k^j\,l^i\ ,\label{19}\\
N^{ij}&=&m^i\,k^j-m^j\,k^i\ .\label{20}
\end{eqnarray}
Each of these bivectors is self--dual in the sense that ${}^*L^{ij}=i\,L^{ij}$, ${}^*M^{ij}=i\,M^{ij}$ and ${}^*N^{ij}=i\,N^{ij}$ with $i=\sqrt{-1}$ and ${}^*L^{ij}=\frac{1}{2}\eta^{ijkm}\,L_{km}$ etc. with $\eta_{ijkm}=\sqrt{-g}\,\epsilon_{ijkm}$, $g={\rm det}(g_{ij})$ and $\epsilon_{ijkm}$ the totally skewsymmetric Levi--Civita permutation symbol in four dimensions with $\epsilon_{1234}=1$. Any self--dual complex bivector can be expressed as a linear combination of these three self--dual bivectors. In particular we have from (\ref{13})
\begin{equation}\label{21}
F_{ij}-i{}^*F_{ij}=2\,q^{-1}f_{\zeta}\,N_{ij}\ .
\end{equation}
The covariant vector field $k_i$ and the complex bivector $N_{ij}$ satisfy the important equations, in the current context,
\begin{equation}\label{22}
k_{i;j}=-q^{-1}\bar\beta\,m_i\,k_j-q^{-1}\beta\,\bar m_i\,k_j\ ,
\end{equation}
given in \cite{Hogan:Puetzfeld:2021:1,Hogan:Puetzfeld:2022:book}, and
\begin{equation}\label{23}
N_{ij;k}=-q^{-1}\beta\,M_{ij}\,k_k\ ,
\end{equation}
so that if $\beta=0$ then $k_i$ and $N_{ij}$ are covariantly constant. Using (\ref{23}) we confirm that (\ref{21}) satisfies Maxwell's equations in the form
\begin{equation}\label{24}
(F^{ij}-i{}^*F^{ij})_{;j}=-2\,q^{-1}f_{\zeta\bar\zeta}\,k^i=0\ .
\end{equation}
It also follows from (\ref{21}) that
\begin{equation}\label{25}
(F_{ij}-i{}^*F_{ij})\,k^j=0\ ,
\end{equation}
indicating that we have here a purely radiative Maxwell field with propagation direction $k^i$ in space--time. Next we can write (\ref{11}) in the form
\begin{equation}\label{26}
C^{ijkm}-i{}^*C^{ijkm}=-2\,q^{-1}H_{\zeta\zeta}\,N^{ij}\,N^{km}\ .
\end{equation}
Using (\ref{23}) we obtain from this
\begin{equation}\label{27}
(C^{ijkm}-i{}^*C^{ijkm})_{;m}=2\,q^{-1}H_{\zeta\zeta\bar\zeta}\,N^{ij}\,k^k\ .
\end{equation}
The Bianchi identities and the field equations (\ref{1}) result in
\begin{equation}\label{28}
C^{ijkm}{}_{;m}=\frac{1}{2}\kappa\,(E^{ki;j}-E^{kj;i}),
\end{equation}
with
\begin{equation}\label{28_1}
E^{ki}=2\,q^{-2}f_{\zeta}\,f_{\bar\zeta}\,k^k\,k^i\ .
\end{equation}
With the use of (\ref{22}) we find that
\begin{eqnarray}
C^{ijkm}{}_{;m}&=&-\kappa\,q^{-1}(q^{-1}f_{\zeta}\,f_{\bar\zeta})_{\zeta}\,N^{ij}\,k^k \nonumber \\
&&-\kappa\,q^{-1}(q^{-1}f_{\zeta}\,f_{\bar\zeta})_{\bar\zeta}\,\bar N^{ij}\,k^k\ .\label{29}
\end{eqnarray}
Since ${}^*N^{ij}=i\,N^{ij}$ we have
\begin{eqnarray}
{}^*C^{ijkm}{}_{;m}&=&-\kappa\,i\,q^{-1}(q^{-1}f_{\zeta}\,f_{\bar\zeta})_{\zeta}\,N^{ij}\,k^k\nonumber\\
&&+\kappa\,i\,q^{-1}(q^{-1}f_{\zeta}\,f_{\bar\zeta})_{\bar\zeta}\,\bar N^{ij}\,k^k\ .\label{30}
\end{eqnarray}
Hence the left hand side of (\ref{27}) can be written
\begin{equation}\label{31}
(C^{ijkm}-i{}^*C^{ijkm})_{;m}=-2\,\kappa\,q^{-1}(q^{-1}f_{\zeta}\,f_{\bar\zeta})_{\zeta}\,N^{ij}\,k^k\ .
\end{equation}
Therefore (\ref{27}) is automatically satisfied since it reduces to the derivative with respect to $\zeta$ of the field equation (\ref{16}). Finally we see from (\ref{26}) that
\begin{equation}\label{32}
(C^{ijkm}-i{}^*C^{ijkm})\,k_m=0\ ,
\end{equation}
indicating pure gravitational radiation with propagation direction $k^i$ in space--time.

\section{A Gravito--electromagnetic Searchlight}\label{sec:3}

\begin{figure}
\includegraphics[width=0.25\textwidth]{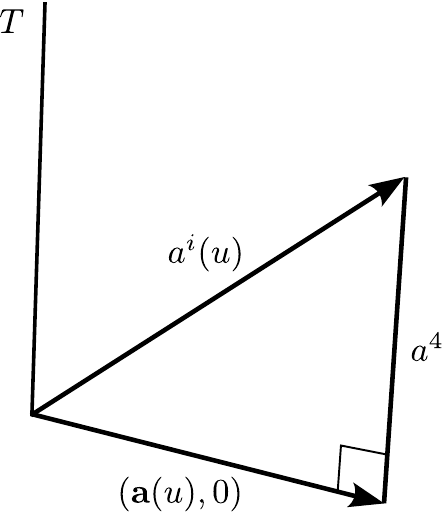}
\caption{\label{fig_1} $a^i(u)$ is resolved into $({\bf a}(u), 0)$ in the observer's space and a vector of magnitude $a^4(u)$ parallel to the T-axis. The 3-velocity of the waves is ${\bf v}(u) = (a^4)^{-1} {\bf a}$  in the observer's space.}
\end{figure}

In the geometrical construction of (\ref{7}), described in \cite{Hogan:Puetzfeld:2021:1,Hogan:Puetzfeld:2022:book}, the function $q(\zeta, \bar\zeta, u)$ given in (\ref{8}) emerges naturally. We can use it to provide a simple explicit example of the Einstein--Maxwell fields described above by taking $f=g_0\,q$ in the potential 1--form (\ref{12}) with $g_0$ a real constant . Then the Maxwell field $F_{ij}=0$ except for $F_{14}=g_0\,\bar\beta$ and $F_{24}=g_o\,\beta$ while (\ref{14}) is satisfied. The Einstein--Maxwell field equation (\ref{16}) now reads
\begin{equation}\label{33}
H_{\zeta\bar\zeta}=-\kappa\,g_0^2\,q^{-1}|\beta|^2\ ,
\end{equation} 
and we solve this with $H=-\kappa\,g_0^2\,q\,\log q$. Now the Weyl conformal curvature tensor $C_{ijkm}=0$ except for $C_{1414}=2\,g_0^2\,\bar\beta^2$ and $C_{2424}=2\,g_0^2\,\beta^2$. In this case (\ref{21}) and (\ref{26}) read
\begin{equation}\label{33'}
F_{ij}-i{}^*F_{ij}=2\,g_0\,q^{-1}\bar\beta\,N_{ij}\ ,
\end{equation}
and 
\begin{equation}\label{34'}
C_{ijkm}-i{}^*C_{ijkm}=2\,\kappa\,g_0^2\,q^{-2}\bar\beta^2N_{ij}\,N_{km}\ ,
\end{equation}
and so we have the Weyl double copy (cf. \cite{Godazgar:etal:2021:1})
\begin{equation}\label{34''}
C_{ijkm}-i{}^*C_{ijkm}=\frac{1}{2}\kappa\,(F_{ij}-i{}^*F_{ij})(F_{km}-i{}^*F_{km})\ .
\end{equation}
The potential 1--form (\ref{12}) now reads
\begin{equation}\label{34}
A=g_0\,q\,du\ ,
\end{equation}
and the line element (\ref{7}) reads
\begin{equation}\label{35}
ds^2=2\,|d\zeta-\beta(u)\,v\,du|^2+2\,q\,du\,(dv-\kappa\,g_0^2\,q\,\log q\,du)\ .
\end{equation}
This gravito--electromagnetic searchlight can be written in a more surveyable form by introducing coordinates $X^i=(X, Y, Z, T)$ with $i=1, 2, 3, 4$ via the transformations \cite{Hogan:Puetzfeld:2021:1,Hogan:Puetzfeld:2022:book}
\begin{eqnarray}
X+i\,Y&=&\sqrt{2}\,\zeta-\sqrt{2}\,l(u)\,v,\label{36}\\
Z+T&=&2\{\bar l(u)\,\zeta+l(u)\,\bar\zeta\}-2\,l(u)\,\bar l(u)\,v+n(u),\label{37}\\
Z-T&=&v,\label{38}
\end{eqnarray}
with $dl/du=\beta$ and $dn/du=\gamma$. Applying these transformations to the line element (\ref{35}) results in 
\begin{eqnarray}
ds^2&=&(dX)^2+(dY)^2+(dZ)^2-(dT)^2 \nonumber\\
&&-2\,\kappa\,g_0^2\,q^2\log q\,du^2\ \nonumber\\
&=&\eta_{ij}\,dX^i\,dX^j-2\,\kappa\,g_0^2\,q^2\log q\,du^2\ ,\label{39}
\end{eqnarray}
where $\eta_{ij}={\rm diag}(1, 1, 1, -1)$ is the Minkowskian metric tensor in rectangular Cartesian coordinates and time. Solving (\ref{36})--(\ref{38}) for $(\zeta, \bar\zeta, v, u)$ in terms of $X^i$ is straightforward except for the equation giving the coordinate $u$ as a function of $X^i$. This equation is obtained by substituting (\ref{36}) and (\ref{38}) into (\ref{37}) to arrive at
\begin{equation}\label{40}
\sqrt{2}\,(\bar l+l)\,X+i\,\sqrt{2}\,(\bar l-l)\,Y+(2\,l\,\bar l-1)\,Z-(2\,l\,\bar l+1)\,T+n=0\ ,
\end{equation}
giving $u(X^i)$ implicitly. Defining 
\begin{equation}\label{41}
a^i(u)=\left (\sqrt{2}\,(\bar l+l), i\,\sqrt{2}\,(\bar l-l), 2\,l\,\bar l-1, 2\,l\,\bar l+1\right )\ ,
\end{equation}
we have
\begin{equation}\label{42}
\eta_{ij}\,a^i\,a^j=a_j\,a^j=0\ ,
\end{equation}
and (\ref{40}) takes the form
\begin{equation}\label{43}
a_i(u)\,X^i+n(u)=0\ .
\end{equation}
We thus see that the hypersurfaces $u={\rm constant}$ are null hyperplanes in Minkowskian space--time. These hyperplanes intersect provided $\dot a^i=da^i/du\neq 0\ (\Leftrightarrow\ \beta\neq 0)$. From (\ref{43}) we have 
\begin{equation}\label{44}
du=u_{,i}\,dX^i=-W\,a_i\,dX^i\ \ \ {\rm with}\ \ \ W=(\dot a_j\,X^j+\gamma)^{-1}\ .
\end{equation}
With $q(\zeta, \bar\zeta, u)$ given by (\ref{8}) we find, using (\ref{36}) and (\ref{38}),
\begin{eqnarray}
2\,q&=&\sqrt{2}\,(\bar\beta+\beta)\,X+i\,\sqrt{2}\,(\bar\beta-\beta)\,Y \nonumber \\
&&+2\,(\bar\beta\,l+\beta\,\bar l)(Z-T)+\gamma\ \nonumber\\
&=&\dot a^1\,X+\dot a^2\,Y+\dot a^3\,Z-\dot a^4\,T+\gamma\ ,\label{45}
\end{eqnarray}
so that
\begin{equation}\label{46}
2\,q=\dot a_j\,X^j+\gamma=W^{-1}\ .
\end{equation}
Hence we have
\begin{equation}\label{47}
q\,du=-\frac{1}{2}\,a_i\,dX^i\ ,
\end{equation}
and so the line element (\ref{39}) written entirely in the coordinates $X^i$ reads
\begin{equation}\label{48}
ds^2=(\eta_{ij}+\frac{1}{2}\kappa\,g_0^2\,\log W\,a_i\,a_j)\,dX^i\,dX^j\ .
\end{equation}

We note in passing that $W(X^i)$ given by (\ref{44}) is, with respect to the Minkowskian metric, a wave function and its gradient is orthogonal to $a^i$. With a comma denoting partial differentiation with respect to $X^i$ the first and second derivatives of $W$ are
\begin{equation}\label{49}
W_{,i}=(\ddot a_j\,X^j+\dot\gamma)\,W^3\,a_i-W^2\,\dot a_i\ ,
\end{equation}
and
\begin{eqnarray}
W_{,ij}&=&W^3(a_i\,\ddot a_j+\ddot a_i\,a_j+2\,\dot a_i\,\dot a_j) \nonumber \\
&&-3\,(\ddot a_k\,X^k+\dot\gamma)\,W^4(a_i\,\dot a_j+\dot a_i\,a_j)\nonumber\\
&&+\biggl\{3\,(\ddot a_k\,X^k+\dot\gamma)^2W^5-(\dddot a_k\,X^k+\ddot\gamma)W^4\biggr\}a_i\,a_j\ .\nonumber \\\label{50}
\end{eqnarray}
Hence 
\begin{equation}\label{51}
W_{,i}\,a^i=0\ ,\ W_{,i}\,\dot a^i=-W^2\dot a_i\,\dot a^i=W^2a_i\,\ddot a^i\ ,
\end{equation}
so that, in particular, the gradient of $W$ is orthogonal to $a^i$ with respect to the Minkowskian metric, and
\begin{equation}\label{52}
\Box W=\eta^{ij}\,W_{,ij}=2\,W^3(a_i\,\ddot a^i+\dot a_i\,\dot a^i)=0\ ,
\end{equation}
so that $W$ is a Minkowskian wave function.

In coordinates $X^i$ the gravito--electromagnetic searchlight has metric tensor of Kerr--Schild form
\begin{eqnarray}
g_{ij}&=&\eta_{ij}+\frac{1}{2}\kappa\,g_0^2\,\log W\,a_i(u)\,a_j(u)\nonumber \\ 
\Rightarrow\ \ g^{ij}&=&\eta^{ij}-\frac{1}{2}\kappa\,g_0^2\log W\,a^i\,a^j\ ,\label{53}
\end{eqnarray}
with $a^i=\eta^{ij}\,a_j=g^{ij}\,a_j$. $W$ is given by (\ref{44}) and $u(X^i)$ implicitly by (\ref{43}). The electromagnetic 4--potential is 
\begin{equation}\label{54}
A^i=-\frac{1}{2}\,g_0\,a^i(u)\ .
\end{equation}
Calculating directly with these we see that the Maxwell field is
\begin{equation}\label{55}
F^{ij}=\frac{1}{2}\,g_0\,W\,(\dot a^i\,a^j-\dot a^j\,a^i)\ \ \ \Rightarrow\ F^{ij}\,a_j=0\ .
\end{equation} 
Denoting by a semicolon covariant differentiation with respect to the Riemannian connection calculated with the metric tensor (\ref{53}) and noting from the metric form that $g={\rm det}(g_{ij})={\rm det}(\eta_{ij})=-1$ we satisfy the gauge condition
\begin{equation}\label{56}
A^i{}_{;i}=\frac{1}{\sqrt{-g}}(\sqrt{-g}\,A^i)_{,i}=A^i{}_{,i}=\frac{1}{2}\,g_0\,W\,\dot a^i\,a_i=0\ .
\end{equation}
Also we confirm that Maxwell's equations are satisfied since
\begin{eqnarray}
F^{ij}{}_{;j}&=&\frac{1}{\sqrt{-g}}(\sqrt{-g}\,F^{ij})_{,j}=F^{ij}{}_{,j}\nonumber\\
&=&\frac{1}{2}\,g_0\,W_{,j}\,a^j\,\dot a^i+\frac{1}{2}\,g_0\,(W^2a_j\,\ddot a^j-W_{,j}\,\dot a^j)\,a^i=0\ ,\nonumber \\ \label{57}
\end{eqnarray}
with the final equality following from (\ref{51}). The electromagnetic energy--momentum tensor is given by
\begin{equation}\label{58}
E_{ij}=\frac{1}{4}\,g_0^2\,W^2\dot a_p\,\dot a^p\,a_i\,a_j\ .
\end{equation}
A calculation of the Riemann curvature tensor components starting with the metric tensor (\ref{53}) results in
\begin{equation}\label{59}
R_{ijkm}=\frac{1}{4}\kappa\,g_0^2\,W^2(\dot a_i\,a_j-a_i\,\dot a_j)(\dot a_k\,a_m-a_k\,\dot a_m)\ ,
\end{equation}
from which the Ricci tensor components are found to be
\begin{equation}\label{60}
R_{jk}=g^{im}\,R_{ijkm}=-\frac{1}{4}\kappa\,g_0^2\,W^2\dot a_m\,\dot a^m\,a_j\,a_k\ .
\end{equation}
The equations (\ref{58}) and (\ref{60}) confirm that the Einstein--Maxwell field equations $R_{ij}=-\kappa\,E_{ij}$ are satisfied. With $A^i$ given by (\ref{54}), satisfying the gauge condition (\ref{56}), and the Ricci tensor (\ref{60}), it is straightforward to check that the Einstein--Maxwell field equations in the form of (\ref{4}) are satisfied. Finally the Weyl conformal curvature tensor components in coordinates $X^i$ are
\begin{eqnarray}
C_{ijkm}&=&\frac{1}{4}\kappa\,g_0^2\,W^2\{S_{ik}\,a_j\,a_m+S_{jm}\,a_i\,a_k\nonumber \\
&&-S_{im}\,a_j\,a_k-S_{jk}\,a_i\,a_m\}\ ,\label{61}
\end{eqnarray}
with
\begin{equation}\label{62}
S_{ij}=\dot a_i\,\dot a_j-\frac{1}{2}\,\dot a_p\,\dot a^p\,\eta_{ij}=S_{ji}\ \ \Rightarrow\ \ S_{ij}\,a^j=-\frac{1}{2}\,\dot a_p\,\dot a^p\,a_i\ .
\end{equation}
Consequently we have the radiative property (the partner for (\ref{55}))
\begin{equation}\label{63}
C_{ijkm}\,a^m=0\ .
\end{equation}

\section{Fields on Minkowskian Space--Time}\label{sec:4}

\begin{figure}
\includegraphics[width=0.3\textwidth]{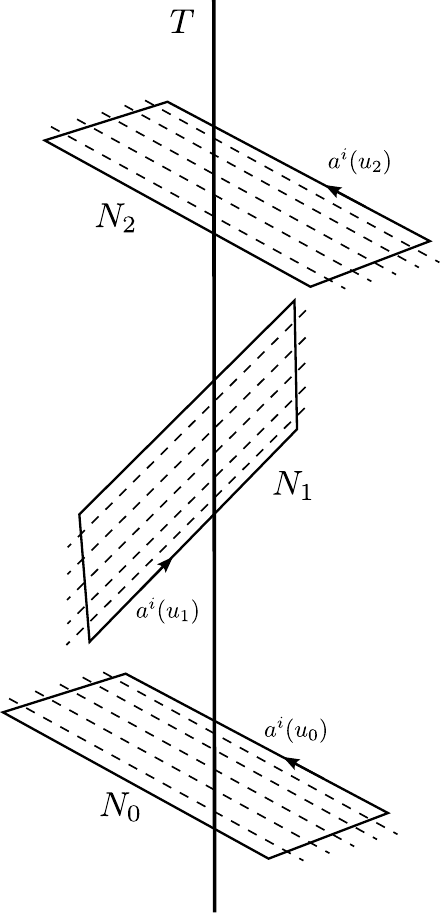}
\caption{\label{fig_2} The intersecting null hyperplane histories $N_0(u_0), N_1(u_1), N_2(u_2)$ of some plane wave fronts as the gravito-electromagnetic searchlight sweeps across the sky with angular velocity $|\omega| = W (\dot{a}_m \dot{a}^m)^{1/2}$ in the observer's space. The generators of the null hyperplanes (the integral curves of $a^i(u)$ in each case) are indicated.}
\end{figure}

It follows from (\ref{57}) that $F^{ij}$ given by (\ref{55}) is not only a Maxwell field on the space--time with line element (\ref{53}) but is also a Maxwell field on Minkowskian space--time with metric tensor components $\eta_{ij}$ in coordinates $X^i=(X, Y, Z, T)$. This Maxwell field on Minkowskian space--time determines an electric 3--vector ${\bf E}=(E^1, E^2, E^3)$ and a magnetic 3--vector ${\bf B}=(B^1, B^2, B^3)$ given respectively by
\begin{equation}\label{64}
E^{\alpha}=F^{\alpha4}\ ,\ B^{\alpha}={}^*F^{\alpha4}\ ,
\end{equation}
with Greek indices taking values 1, 2, 3. With $a^i(u)$ in (\ref{41}) we have from (\ref{55}):
\begin{equation}\label{65}
{\bf B}+i\,{\bf E}=2\,i\,g_0\,W\,(2\,l\,\bar l+1)\,\bar\beta\,{\bf m} ,
\end{equation}
with the 3--vector ${\bf m}$ given by
\begin{equation}\label{66}
{\bf m}=\left (\frac{1-2\,l^2}{\sqrt{2}\,(2\,l\,\bar l+1)}, \frac{i\,(1+2\,l^2)}{\sqrt{2}\,(2\,l\,\bar l+1)}, \frac{2\,l}{2\,l\,\bar l+1}\right )\ .
\end{equation}
The formula (\ref{65}) with (\ref{66}) is a special case of a formula satisfied by plane fronted electromagnetic waves in general given in \cite{Hogan:Puetzfeld:2021:1,Hogan:Puetzfeld:2022:book}. The histories in Minkowskian space--time of the plane fronted waves are the null hyperplanes $u(X^i)={\rm constant}$ given by (\ref{43}). The \emph{wave velocity} (see \cite{Synge:1956:book}) has components
\begin{equation}\label{67}
v^{\alpha}=-\frac{u_{,4}\,u_{,\alpha}}{u_{,\beta}\,u_{,\beta}}\ ,
\end{equation}
which, as a consequence of (\ref{44}) reads (see Fig.\ \ref{fig_1})
\begin{equation}\label{68}
{\bf v}=(a^4)^{-1}{\bf a}\ ,
\end{equation}
with $a^i=({\bf a}, a^4)$ by (\ref{41}). The plane fronted waves, of course, travel with the speed of light so that ${\bf v}\cdot{\bf v}=1$. With $u_{,4}=-W\,a_4=W\,a^4$ we find that
\begin{equation}\label{69}
\frac{\partial {\bf v}}{\partial T}=\frac{\partial {\bf v}}{\partial u}\,u_{,4}=2\,W\,(\bar\beta\,{\bf m}+\beta\,\bar{\bf m})\ ,
\end{equation}
with $\bar{\bf m}$ the complex conjugate of ${\bf m}$. Among the useful properties of the complex 3--vector ${\bf m}$ (${\bf m}\cdot\bar{\bf m}=1$,\ ${\bf m}\cdot{\bf m}=0=\bar{\bf m}\cdot\bar{\bf m}$ and ${\bf m}\times\bar{\bf m}=i\,{\bf v}$ ) are the vector products 
\begin{equation}\label{70}
{\bf m}=i\,{\bf m}\times{\bf v}\ \ \ \ {\rm and}\ \ \ \ \bar{\bf m}=-i\,\bar{\bf m}\times{\bf v}\ .
\end{equation}
Consequently it follows from (\ref{65}) that $|{\bf E}|^2=|{\bf B}|^2$, ${\bf E}\cdot{\bf B}=0$, ${\bf E}\times{\bf B}=|{\bf E}|^2{\bf v}$ and ${\bf E}, {\bf B}, {\bf v}$ form a right handed orthogonal triad. In addition (\ref{69}) can be written
\begin{equation}\label{71}
\frac{\partial {\bf v}}{\partial T}={\bf \omega}\times{\bf v}\ \ \ \ {\rm with}\ \ \ \ {\bf\omega}=2\,i\,W\,(\bar\beta\,{\bf m}-\beta\,\bar{\bf m})\ ,
\end{equation}
so that $|{\bf\omega}|^2={\bf\omega}\cdot{\bf\omega}=8\,W^2|\beta|^2$. Hence provided $\beta\neq 0$ (so that the null hyperplanes (\ref{43}) with $u={\rm constant}$ \emph{intersect} and the wave fronts \emph{collide}) ${\bf\omega}$ is the angular velocity with which the searchlight sweeps across the sky as described by Pirani \cite{Pirani:1965} (see Fig.\ \ref{fig_2}). If $\beta=0$ there is no searchlight. The searchlight we are considering here can also project gravitational radiation and is then a gravito--electromagnetic searchlight. We achieve this in the present context by considering (\ref{53}) as a perturbation of Minkowskian space--time written as
\begin{equation}\label{72}
g_{ij}=\eta_{ij}+\gamma_{ij}\ ,
\end{equation}
where
\begin{equation}\label{73}
\gamma_{ij}=\frac{1}{2}\kappa\,g_0^2\,\log W\,a_i(u)\,a_j(u)\ ,
\end{equation}
is a tensor field on Minkowskian space--time. Using (\ref{42}), (\ref{44}) and (\ref{49}) we find that $\gamma_{ij}$ satisfies the gauge condition
\begin{equation}\label{74}
\eta^{jk}\,\gamma_{ij,k}=0\ .
\end{equation}
The linearized Riemann curvature tensor components are
\begin{equation}\label{75}
L_{ijkm}=\frac{1}{2}(\gamma_{im,jk}+\gamma_{jk,im}-\gamma_{ik,jm}-\gamma_{jm,ik})\ ,
\end{equation}
and so, making use of (\ref{50}), the components of the linearized Ricci tensor are
\begin{eqnarray}
L_{jk}&=&\eta^{im}\,L_{ijkm}=\frac{1}{2}\eta^{im}\,\gamma_{jk,im}\nonumber \\
&=&-\frac{1}{4}\kappa\,g_0^2W^2\dot a_m\,\dot a^m\,a_j\,a_k=-\kappa\,E_{jk}\ ,\label{76}
\end{eqnarray}
with $E_{jk}$ given by (\ref{58}). On account of the Weyl double copy (\ref{34''}) the linearized Weyl tensor is given in coordinates $X^i$ by
\begin{equation}\label{77}
C^{ijkm}-i{}^*C^{ijkm}=\frac{1}{2}\kappa\,(F^{ij}-i{}^*F^{ij})(F^{km}-i{}^*F^{km})\ .
\end{equation}
In terms of the electric part of the linearized Weyl tensor $E^{\alpha\beta}=C^{\alpha4\beta4}$ and the magnetic part $B^{\alpha\beta}={}^*C^{\alpha4\beta4}$ we obtain from this the gravitational partner for (\ref{65}) namely
\begin{equation}\label{78}
B^{\alpha\beta}+i\,E^{\alpha\beta}=2\,i\,\kappa\,g_0^2W^2\bar\beta^2(2\,l\,\bar l+1)^2m^{\alpha}\,m^{\beta}\ ,\end{equation} with $(m^{\alpha})={\bf m}$. The plane fronted gravitational waves described by this tensor share their wave fronts with the electromagnetic waves above and their wave fronts undergo the rotation implied by (\ref{69}) making the searchlight a gravito--electromagnetic searchlight. Using (\ref{65}) and (\ref{78}) we deduce that
\begin{equation}\label{79}
E^{\alpha\beta}=\frac{1}{2}\kappa\,(E^{\alpha}\,E^{\beta}-B^{\alpha}\,B^{\beta})=E^{\beta\alpha}\ \ \ \Rightarrow\ \ \ E^{\alpha\alpha}=0\ ,
\end{equation}
and
\begin{equation}\label{79_1}
B^{\alpha\beta}=\frac{1}{2}\kappa\,(E^{\alpha}\,B^{\beta}+E^{\beta}\,B^{\alpha})=B^{\beta\alpha}\ \ \ \Rightarrow\ \ \ B^{\alpha\alpha}=0\ .
\end{equation}
Using the algebraic relations following (\ref{70}) above involving ${\bf E}, {\bf B}$ and ${\bf v}$ we obtain the corresponding relations involving $E^{\alpha\beta}$, $B^{\alpha\beta}$ and ${\bf v}$:
\begin{eqnarray}
E^{\alpha\beta}\,E^{\beta\sigma}&=&B^{\alpha\beta}\,B^{\beta\sigma}\ , \label{80} \\
E^{\alpha\beta}\,B^{\beta\sigma}+B^{\alpha\beta}\,E^{\beta\sigma}&=&0\ ,\label{81}
\end{eqnarray}
and
\begin{equation}\label{82}
\epsilon_{\sigma\rho\lambda}\,E^{\alpha\rho}\,B^{\beta\lambda}=E^{\alpha\rho}\,E^{\rho\beta}\,v^{\sigma}=B^{\alpha\rho}\,B^{\rho\beta}\,v^{\sigma}\ ,
\end{equation}
where $\epsilon_{\sigma\rho\lambda}$ is the three dimensional Levi--Civita permutation symbol.

\section{Discussion}\label{sec:5}

The electromagnetic field and the gravitational field of our explicit model of a gravito--electromagnetic searchlight are given by (\ref{33'}) and (\ref{34'}) respectively. Their existence depends upon the non--vanishing of the complex valued function $\beta(u)$. This in turn means that the plane fronted waves, both electromagnetic and gravitational, must have colliding wave fronts (see \cite{Hogan:Puetzfeld:2021:1,Hogan:Puetzfeld:2022:book}). A useful property of these fields is given by the so--called Weyl double copy in the form of (\ref{34''}). In the context of tensor fields on Minkowskian space--time in section 4 this leads to simple relationships (\ref{80}) and (\ref{81}) between the electric and magnetic 3--vectors of the electromagnetic field and the so--called electric and magnetic parts of the gravitational field. In our treatment of the gravito--electromagnetic searchlight in coordinates $X^i$ in section 3 an interesting function $W$, which appears first in (\ref{44}), plays a key role. Its most important attribute is that it is a wave function on Minkowskian space--time which we have utilized in confirming that (\ref{4}) is satisfied by the searchlight field and also in deriving (\ref{76}). A more general scenario than that considered here emerges from the asymptotic limit (in the manner of \cite{Hogan:Puetzfeld:2021:2}) of the Li\'enard--Wiechert electromagnetic field. The limiting case is algebraically general but contains a radiation part (just as the Li\'enard--Wiechert field does). A challenging open question is to find a corresponding gravito--electromagnetic model in Einstein--Maxwell theory. 

\bibliographystyle{unsrtnat}
\bibliography{collplanesearchlight_bibliography}

\begin{thebibliography}{10}
\providecommand{\natexlab}[1]{#1}
\providecommand{\url}[1]{\texttt{#1}}
\expandafter\ifx\csname urlstyle\endcsname\relax
  \providecommand{\doi}[1]{doi: #1}\else
  \providecommand{\doi}{doi: \begingroup \urlstyle{rm}\Url}\fi

\bibitem[{Hogan} and {Puetzfeld}(2021{\natexlab{a}})]{Hogan:Puetzfeld:2021:1}
P.~A. {Hogan} and D.~{Puetzfeld}.
\newblock {Gravitational waves with colliding or noncolliding wave fronts}.
\newblock \emph{Phys. Rev. D}, 103:\penalty0 124064, 2021{\natexlab{a}}.
\newblock \doi{10.1103/PhysRevD.103.124064}.

\bibitem[{Hogan} and {Puetzfeld}(2022)]{Hogan:Puetzfeld:2022:book}
P.~A. {Hogan} and D.~{Puetzfeld}.
\newblock \emph{{Exact Space--Time Models of Gravitational Waves}}.
\newblock Springer, Cham, 2022.
\newblock ISBN 978-3-031-16825-3.

\bibitem[{Kundt}(1961)]{Kundt:1961}
W.~{Kundt}.
\newblock {The plane fronted gravitational waves}.
\newblock \emph{Z. Phys.}, 163:\penalty0 77, 1961.
\newblock \doi{10.1007/BF01328918}.

\bibitem[{Ehlers} and {Kundt}(1962)]{Ehlers:Kundt:1962}
J.~{Ehlers} and W.~{Kundt}.
\newblock {Exact solutions of the gravitational field equations}.
\newblock \emph{In ``Gravitation: An Introduction to Current Research'', edited
  by L.\ Witten (John Wiley and Sons, Inc., New York)}, page~49, 1962.

\bibitem[{Stephani} et~al.(2003){Stephani}, {Kramer}, {MacCallum},
  {Hoenselaers}, and {Herlt}]{Stephani:etal:2003:book}
H.~{Stephani}, D.~{Kramer}, M.~A.~H. {MacCallum}, C.~{Hoenselaers}, and
  E.~{Herlt}.
\newblock \emph{{Exact Solutions of Einstein's Field Equations}}.
\newblock 2nd. Edn., Cambridge University Press, Cambridge, 2003.
\newblock ISBN 978-0-511-53518-5.

\bibitem[{Hogan} and {Puetzfeld}(2021{\natexlab{b}})]{Hogan:Puetzfeld:2021:2}
P.~A. {Hogan} and D.~{Puetzfeld}.
\newblock {Plane fronted limit of spherical electro-magnetic and gravitational
  waves}.
\newblock \emph{Phys. Rev. D}, 104:\penalty0 124081, 2021{\natexlab{b}}.
\newblock \doi{10.1103/PhysRevD.104.124081}.

\bibitem[{Synge}(1960)]{Synge:1960:book}
J.~L. {Synge}.
\newblock \emph{{Relativity: The general theory}}.
\newblock North-Holland, Amsterdam, 1960.

\bibitem[{Godazgar} et~al.(2021){Godazgar}, {Godazgar}, {Veiga}, and
  {Pope}]{Godazgar:etal:2021:1}
H.~{Godazgar}, M.~{Godazgar}, R.~{Monteiro} D.~P. {Veiga}, and C.~N. {Pope}.
\newblock {Weyl double copy for gravitational waves}.
\newblock \emph{Phys. Rev. Lett}, 126:\penalty0 101103, 2021.
\newblock \doi{10.1103/PhysRevLett.126.101103}.

\bibitem[{Synge}(1956)]{Synge:1956:book}
J.~L. {Synge}.
\newblock \emph{{Relativity: The special theory}}.
\newblock North--Holland Publishing Company, Amsterdam, 1956.

\bibitem[{Pirani}(1965)]{Pirani:1965}
F.~A.~E. {Pirani}.
\newblock {Introduction to gravitational radiation theory}.
\newblock \emph{In ``Lectures on General Relativity'', Brandeis Lectures
  (Prentice--Hall, New Jersey)}, 1:\penalty0 249, 1965.

\end{thebibliography}
\end{document}